\documentclass[12pt,apj]{emulateapj}

\renewcommand{\deg}{\mbox{$^{\circ}$}}

\def\deg{\ifmmode^\circ\else$^\circ$\fi}    

\def\hper{\ifmmode \rlap.^{h}\else $\rlap{.}^h$\fi} 
\def\sper{\ifmmode \rlap.^{s}\else $\rlap{.}^s$\fi}    

\def\deg{${}^\circ$}

\def\rr{\hbox{\em r\/}}
\def\ii{\hbox{\em i\/}}

\def\rmi{\hbox{\em r--i\/}}

\def\today{\number\year\space \ifcase\month\or
  January\or February\or March\or April\or May\or June\or
  July\or August\or September\or October\or November\or December\fi
  \space\number\day}
\def\now{\number\year\space \ifcase\month\or
  January\or February\or March\or April\or May\or June\or
  July\or August\or September\or October\or November\or December\fi
  \space\number\day .\number\time}

\shortauthors{Di Cecco et al.}

\begin{document}
\title{On the density profile of the globular cluster M92\altaffilmark{1}}

\author{
A. Di Cecco\altaffilmark{2,3}, 
A. Zocchi\altaffilmark{4}, 
A. L. Varri\altaffilmark{4}, 
M. Monelli\altaffilmark{5,6}, 
G. Bertin\altaffilmark{4}, 
G. Bono\altaffilmark{7,3}, 
P. B. Stetson\altaffilmark{8,9}, 
M. Nonino\altaffilmark{10} and
R. Buonanno\altaffilmark{7,2}, 
I. Ferraro\altaffilmark{3}, 
G. Iannicola\altaffilmark{3}, 
A. Kunder\altaffilmark{11}, 
A. R. Walker\altaffilmark{11}
} 

\altaffiltext{1}{Based in part on data obtained from the ST-ECF Science
Archive Facility. This research used the facilities of the Canadian Astronomy
Data Centre operated by the National Research Council of Canada with the
support of the Canadian Space Agency. 
}
\altaffiltext{2}{Agenzia Spaziale Italiana Science Data Center (ASDC), c/o
ESRIN, via G. Galilei, 00044 Frascati, Italy}

\altaffiltext{3}{Istituto Nazionale di Astrofisica (INAF) -- Osservatorio Astronomico di Roma, Via Frascati 33, 
00044 Monte Porzio Catone, Italy}

\altaffiltext{4}{Universit\`a degli Studi di Milano, Dipartimento di Fisica, via Celoria 16, 20133 Milano, Italy}

\altaffiltext{5}{Instituto de Astrofisica de Canarias, Calle Via Lactea,
E38200 La Laguna, Tenerife, Spain}

\altaffiltext{6}{Departamento de Astrofisica, Universidad de La Laguna,
Tenerife, Spain}

\altaffiltext{7}{Dipartimento di Fisica, Universit\`a di Roma Tor Vergata, 
via della Ricerca Scientifica 1, 00133 Rome, Italy}

\altaffiltext{8}{Dominion Astrophysical Observatory, Herzberg Institute of
Astrophysics, National Research Council, 5071 West Saanich Road, Victoria, BC V9E 2E7, Canada}

\altaffiltext{9}{Visiting Astronomer, Cerro Tololo Inter-American Observatory, 
National Optical Astronomy Observatories, operated by AURA, Inc., under 
cooperative agreement with the NSF.}

\altaffiltext{10}{INAF--Osservatorio Astronomico di Trieste, via G.B. Tiepolo 11, 40131 Trieste, Italy}

\altaffiltext{11}{Cerro Tololo Inter-American Observatory, National Optical
Astronomy Observatory, Casilla 603, La Serena, Chile}

\date{\centering drafted \today\ / Received / Accepted }

\begin{abstract}
We present new number density and surface brightness profiles for the 
globular cluster M92 (NGC~6341). These profiles are calculated 
from optical images collected with the CCD mosaic camera MegaCam at the 
Canada-France-Hawaii-Telescope and with the Advanced Camera for Surveys 
on the Hubble Space Telescope. The ground-based data were supplemented 
with the Sloan Digital Sky Survey photometric catalog. Special care was 
taken to discriminate candidate cluster stars from field stars and to 
subtract the background contamination from both profiles. 
By examining the contour levels of the number density, we found that 
the stellar distribution becomes clumpy at radial distances larger than 
$\sim$13$\arcmin$, and there is no preferred orientation of contours in space. 
We performed detailed fits of King and Wilson models to the 
observed profiles. The best-fit models underestimate the 
number density inside the core radius. Wilson models better 
represent the observations, in particular in the outermost cluster 
regions: the good global agreement of these models with the observations 
suggests that there is no need to introduce an extra-tidal halo to explain 
the radial distribution of stars at large radial distances. The best-fit 
models for the number density and the surface brightness profiles 
are different, even though they are based on the same observations. 
Additional tests support the evidence that this fact reflects 
the difference in the radial distribution of the stellar tracers that 
determine the observed profiles (main sequence stars for the number 
density, bright evolved stars for the surface brightness).
\end{abstract}


\maketitle

\section{Introduction}

The Galactic globular clusters (GCs) are the oldest ($\sim$11--13~Gyr) 
Galactic systems and have complex internal and external dynamics 
(Gnedin \& Ostriker 1997). The internal dynamics is driven by two-body 
relaxation with a time scale that is typically shorter than their age 
(Meylan \& Heggie 1997). Therefore, the density profile of the innermost 
regions is expected to be well described by King models (1962,1966). 
The outermost regions are characterized by the interaction with external 
tidal forces (Spitzer 1958; Spitzer \& Chavalier 1973; 
Aguilar et al. 1988) and by the evaporation of low-mass stars 
(Spitzer \& Harm 1958). 
These phenomena are expected to produce deviations from the spherical 
King models. By moving beyond the cluster truncation radius ($r_t$) 
the escaping stars can form halos or extended tidal tails.

The first empirical evidence of extra-tidal structures in GCs was found from photographic 
plates (Grillmair et al. 1995; Lehmann \& Scholz 1997; Testa et al. 2000 [T00]; 
Leon et al. 2000). More recently, investigations based on CCD photometry found evidence either 
of tidal tails (Odenkirchen et al. 2001,2003;  Grillmair \& Johnson 2006; Chun et al. 2010; 
Jordi \& Grebel 2010 [JG10]) or of surrounding halos (Lee et al. 2003 [L03]; 
Olszewski et al. 2009; JG10; Correnti et al. 2011) around more than 30 GCs.

An accurate determination of the truncation radius of M92 and of the 
shape of its external regions is still missing, even though several 
investigations have been carried out on this topic (see Table~1). The oldest 
Surface Brightness (SB) profile (obtained from photographic plates and small-format 
CCD images) was provided by Trager et al. (1995, [T95]). The same was later analyzed 
by McLaughlin \& van der Marel (2005, [MLvdM05]). By using photographic plates, 
T00 found evidence of extra-tidal stars at 30$\arcmin$ from the cluster center, 
and provided a surface density map, which shows only marginal evidence for an 
elongation orthogonal to the direction of the Galactic center. More recently, 
L03 by using a mosaic CCD camera, confirmed the occurrence of extra-tidal stars 
and showed that the marginal elongation appears only for the brightest stars. 
JG10 analyzed the Sloan Digital Sky Survey (SDSS) photometric catalog and 
found the same elongated contours detected by T00, even though their data 
do not cover the entire M92 area.

\section{Photometric datasets}

We used both ground-based data collected with the 36 CCD mosaic camera
MegaCam at the Canada-France-Hawaii-Telescope (CFHT) and space data collected
with the Advanced Camera for Surveys (ACS) on the Hubble Space Telescope (HST).
The MegaCam images were collected in the {\it g'},{\it r'},{\it i'},{\it z'} 
bands\footnote{Proposal ID: 2004AC03, PI: J. Clem}.
To increase the radial extent of the sky area covered by our dataset, the
ground-based data (MegaCam images) were supplemented with multiband
({\it g},{\it r},{\it i},{\it z}) data collected by the SDSS 
(Aihara et al. 2011). As a whole the ground-based data 
cover $\sim4^{\circ} \times 4^{\circ}$ around the cluster center, 
but they do not uniformly cover the sky area around M92.

\begin{figure}[!ht]
\begin{center}
\includegraphics[height=0.45\textheight,width=1.0\textwidth]{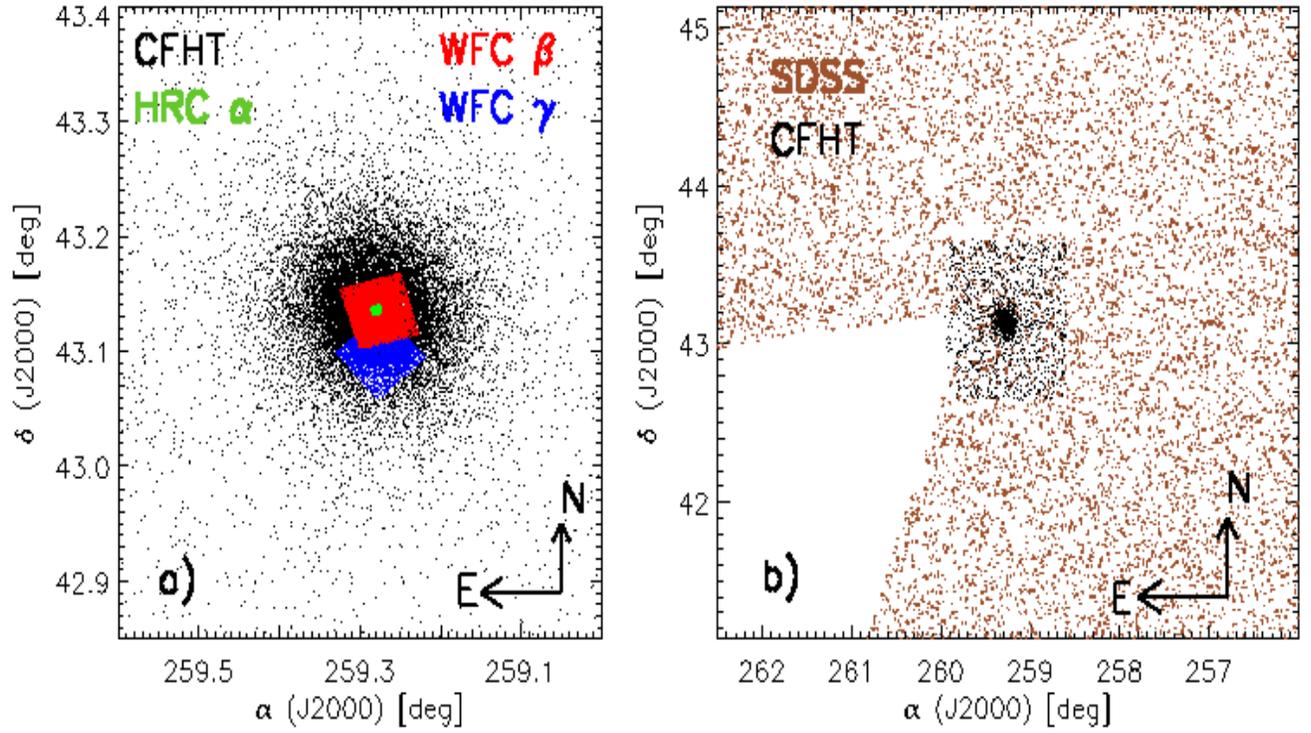}
\vspace*{-0.15truecm}
\caption{Panel a) -- area of the sky across the globular cluster M92 covered by 
the different sets of space images collected with the Advanced Camera for Surveys 
(ACS) on board the HST. 
The field of view of the sky plot is $1^{\circ}\times1^{\circ}$.  
The red and the blue squares show the images collected 
with the Wide Field Channel (WFC, pointings $\beta$, $\gamma$), while the green 
square those collected with the High Resolution Channel (HRC, $\alpha$). 
The black dots display the photometric catalog based on ground-based images 
collected with CFHT. The orientation is shown in the bottom right corner.
Panel b) -- same as Panel a), but for datasets collected with ground-based telescopes, 
namely CFHT (black dots) and SDSS (brown dots). Note that the latter dataset does 
not uniformly cover the area of the sky around M92. 
The field of view of the sky plot is $4^{\circ}\times4^{\circ}$.  
See text for more details.
\label{f1}
}
\end{center}
\end{figure}

The ACS data were collected with three different pointings: {\it pointing}
$\alpha$, High Resolution Channel (HRC), six F435W (t=340 s each) and
155 F555W\footnote{GO-10335, PI: H. Ford} (with exposure times ranging
from 10 to 100 s) images located across the cluster center; {\it pointing}
$\beta$, Wide Field Channel (WFC), three F606W (t=0.5,5,90 s) and three
F814W\footnote{GO-9453, PI: T. Brown} (t=0.5,6,100 s) images located across
the cluster center; {\it pointing} $\gamma$, WFC, three F475W  (t=3,20,40 s)
and three F814W\footnote{GO-10505, PI: C. Gallart} (t=1,10,20 s) images
located 2$\arcmin$ from the cluster center in the South--East direction. 
The reduction and the photometry of the F555W and F814W images have been 
presented and discussed by Di Cecco et al. (2010, [DiC10]).
Panels a) and b) of Fig.~\ref{f1} display the sky coverage of the space-
and ground-based datasets.

To estimate the completeness of the CFHT data we adopted the data of the
pointing $\gamma$ (ACS--WFC). We found that the completeness for $i\le$22 mag
is ∼54\% for 75$\le$R$\le$150 arcsec, ∼82\% for 150$\le$R$\le$200 arcsec, and
complete for larger distances. We applied the above completeness corrections
to the star counts (DiC10). We also estimated the completeness of the SDSS
data by using the CFHT data and we found that for radial distances larger
than ~700 arcsec and $i\le22$ mag they are complete.
For radial distances smaller than $75\arcsec$ we adopted the data of
{\it pointing} $\beta$. These data display a gap of $2.5\arcsec$ between the
two CCDs. To fill this gap we selected two regions of $2.5\arcsec$ at the
edges of the gap and we randomly extracted half of the stars in each of the
two regions. Once the gap was filled, we assumed that the data of
{\it pointing} $\beta$ located between 20$\arcsec$ and 75$\arcsec$ are
complete (DiC10).
The completeness of {\it pointing} $\beta$, for radial distance --$R$-- 
smaller than 11$\arcsec$, was estimated using {\it pointing} $\alpha$.
In this case we found that in the magnitude range between the Main Sequence
Turn Off (MSTO) and {\it F555W}$\sim$21.5 mag the {\it pointing} $\beta$
dataset was complete at the $70\%$ level. The star counts in this cluster
region were corrected accounting for the above completeness correction.

In summary, we are dealing with three different datasets:

{\em 1)}-- ACS--WFC, pointing $\beta$, $R\le1.25\arcmin$. For radial 
distances smaller than $0.18\arcmin$ the completeness was estimated 
using ACS--HRC, pointing $\alpha$. For radial distances 
$0.18\arcmin\le R \le 1.25\arcmin$ we only filled the gaps.

{\em 2)}-- CFHT, $1.25\arcmin < R\le 30\arcmin$. For radial distances
$1.25\arcmin < R\le 3.33\arcmin$ the completeness was estimated using ACS--WFC,
pointing $\gamma$. The comparison of CFHT with pointing $\gamma$ indicates
that the former dataset for $i\le22$ mag  is complete at larger distances.

{\em 3)}-- SDSS, $30\arcmin < R \le 2$\deg. The completeness was estimated
using CFHT data; they are complete for $i\le22$ mag and radial distances larger
than ~700 arcsec. However, they do not uniformly cover the sky region around
the cluster (see panel b) in Fig.~1) and the star counts were accordingly 
corrected.

\begin{figure}[!ht]
\begin{center}
\includegraphics[height=0.55\textheight,width=1.0\textwidth]{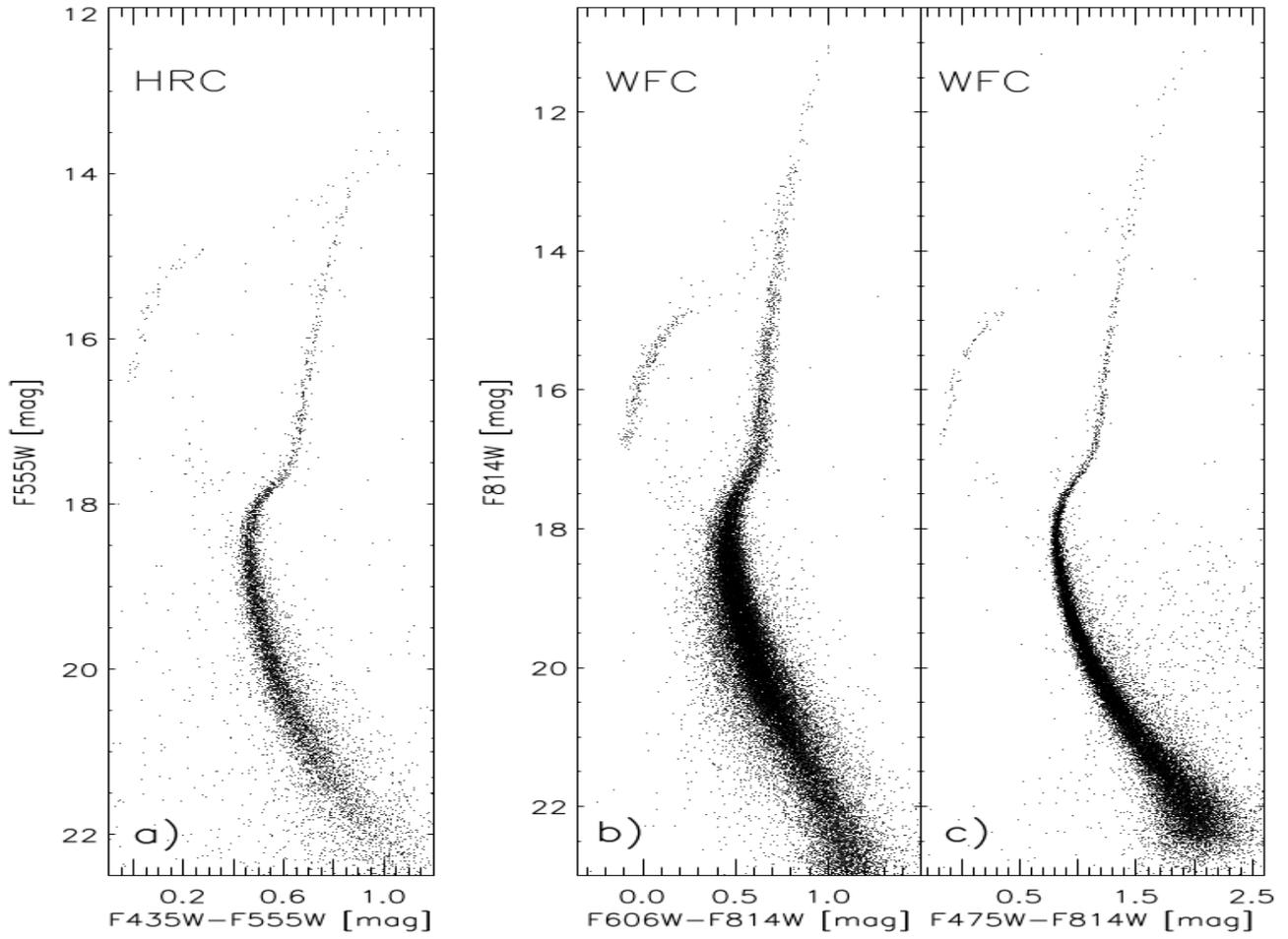}
\vspace*{-0.15truecm}
\caption{CMDs of M92 based on ACS at HST, pointings $\alpha$, $\beta$, and $\gamma$ 
in panels a), b), and c), respectively. Stars plotted in these CMDs were selected 
according to different selection criteria. Data of pointing $\gamma$ are the same 
as in DiC10.
\label{f2}
}
\end{center}
\end{figure}

Panels a), b) and c) of Fig.~\ref{f2} show the color-magnitude diagrams (CMDs)
of {\it pointing} $\alpha$, $\beta$ and $\gamma$. 
%
Stars plotted in these panels were selected according to the {\it sharpness}
($\mid$sh$\mid\leq$1) DAOPHOT index (see, e.g., Stetson 1987, 1994).
These CMDs display well defined sequences in the evolved phases as well as 
along the Main Sequence (MS). The Red Giant Branch (RGB) stars brighter than red Horizontal 
Branch (HB) stars ({\it F555W}$\sim$14 mag) are saturated in {\it pointing} 
$\alpha$.

To provide homogeneous star counts across the entire GC, the {\it i}- and 
the {\it r}-band from the SDSS, as well as the {\it F814W}- and the 
{\it F606W}-band of {\it pointing} $\beta$ were transformed into the {\it i'}- 
and the {\it r'}-band of the MegaCam photometric system. The accuracy of the 
transformations is better than 0.02 mag (DiC10). In the following, we use 
the $g,r,i,z$ bands (without prime) to refer to the CFHT bands.
The reason for the above transformations is threefold: 
{\em a)} \ii- and \rr-band are common to the different datasets; 
{\em b)} the data in these bands have good photometric accuracy 
($\sigma_{\rm r-i}$=0.06 mag, at least three magnitudes fainter than the MSTO);
{\em c)} the \ii-band is minimally affected by saturation problems and it was
adopted to compute the density profiles.

\section{Radial density profile} \label{Sect.3}

The field of view of the CFHT dataset ($1^{\circ}\times1^{\circ}$) and of the
SDSS dataset fully enclose the estimated radial extent of M92 (see Table~1).
In order to constrain the cluster edges, candidate cluster and field stars must 
be distinguished. The method used to identify the two different groups of star 
is described below.

We selected the photometric catalog by using the intrinsic photometric 
error ($\sigma_{\rm{r-i}}\le$0.10 mag), the 
separation\footnote{The separation index quantifies the degree of crowding, 
i.e. the amount of spurious light, due to neighboring stars, affecting the 
magnitude of individual stars (Stetson et al. 2003).} ({\it sep}$\ge$2.5), 
and the distance from the cluster center (10$\arcsec\le R\le$180\arcsec). 
We computed a fiducial line ({\it ridgeline}) in the \ii,\rmi~CMD by means 
of a three-dimensional Hess diagram (Ferraro et al. 2013, in preparation).
Note that the above selection criteria were only applied to estimate the 
ridge line.

\begin{figure}[!ht]
\begin{center}
\includegraphics[height=0.55\textheight,width=1.0\textwidth]{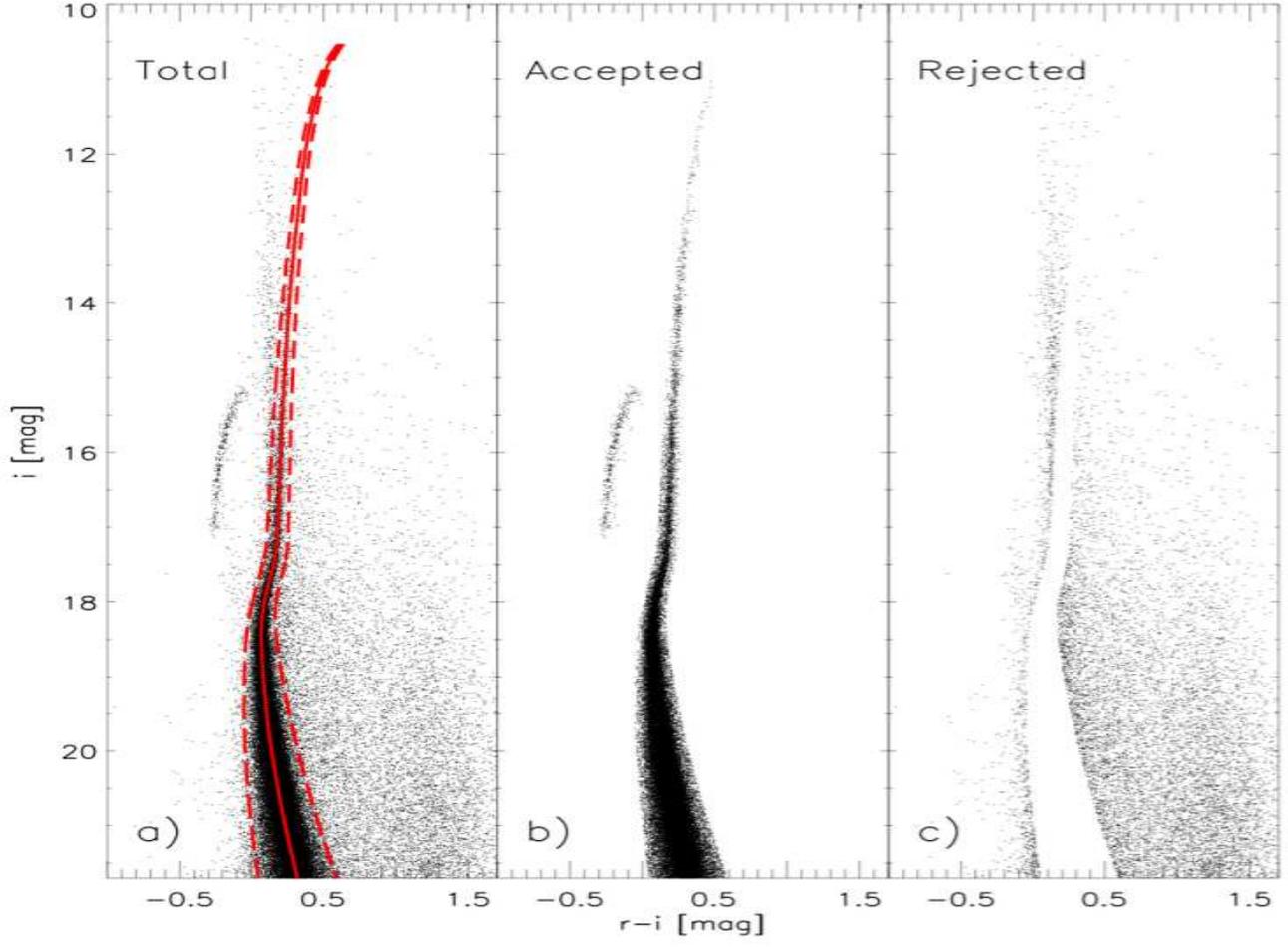}
\vspace*{-0.15truecm}
\caption{Panel a) {\it i,r-i} CMD based on both ground-based (MegaCam at CFHT) 
and space (ACS-WFC at HST) data. Panels b) and c): CMDs for candidate M92 
({\it accepted}) and candidate field ({\it rejected}) stars, respectively. 
See text for more details.
\label{f3}
}
\end{center}
\end{figure}

Panel a) of Fig.~\ref{f3} shows the \ii,\rmi~ CMD for the entire sample of
stars together with the above ridgeline. Stars plotted in this figure were 
selected according to the radial distance and to a very mild photometric error cut
($\sigma_{\rm{r-i}}\le$0.25 mag). Note that in this analysis we only adopted
space and CFHT data; the SDSS data will be discussed in the following.
The solid line shows the computed ridgeline, whereas the dashed ones display
the acceptance region, that is, the region of the CMD where we assume that the
candidate MS and RGB cluster stars lie. The acceptance region is centered on the
ridgeline and the width in color goes from 0.01 mag close to the tip of the RGB
up to 0.30 mag for magnitudes fainter than the MSTO. We cut the acceptance region
at magnitudes brighter than \ii=21.7 mag, because at fainter magnitudes the
photometric error in the color increases and the ridgeline is less well determined.

Panel b) of Fig.~\ref{f3} shows the {\it accepted} M92 stars, that is,
the candidate MS, RGB, and HB stars. The MS and RGB {\it accepted} cluster stars
include some candidate field stars with colors and magnitudes similar to those
of M92. The HB stars, instead, can be easily selected since they are bluer than
the field stars. Panel c) of Fig.~\ref{f3} shows the CMD of candidate field stars
({\it rejected}) with their typical peaks in color around \rmi=0--0.2  and
\rmi=1.2--1.4 mag. The stars located at (\ii,\rmi)$\sim$(17.5,0.0) are
M92 Blue Stragglers. They were not included among {\it accepted} stars, because
the fainter ones partially overlap with field and MSTO stars in the CMD, while
the brighter ones are a minimal fraction of cluster stars.

\begin{figure}[!ht]
\begin{center}
\includegraphics[height=0.55\textheight,width=1.0\textwidth]{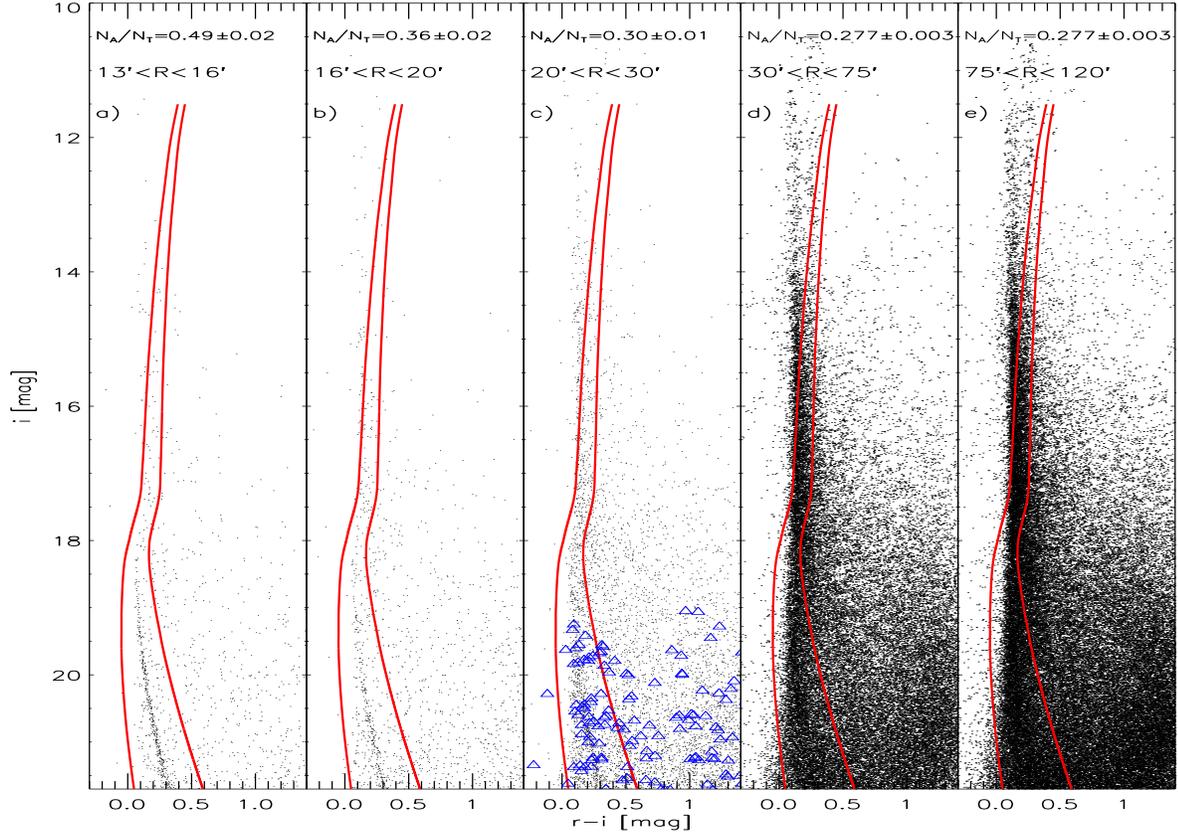}
\vspace*{-0.15truecm}
\caption{Panel a) -- {\it i,r-i} CMD based on MegaCam/CFHT data. The red 
solid lines display the acceptance region defined in Fig.~\ref{f3}. The radial 
extent of the bin and the ratio between candidate cluster stars and 
total number of stars are also labeled.  
Panels b) and c) -- Same as for panel a), but for stars located at larger radial 
distances. The blue triangles plotted in panel c) are the objects located inside 
the candidate galaxy clusters (Wen et al. 2009). See text for more details.   
Panels d) and e) -- same as for panel a), but for stars entirely located inside the 
SDSS dataset.    
\label{f4}
}
\end{center}
\end{figure}

To further constrain the radial extent of candidate cluster stars we investigated
the ratio between the number of accepted stars and the total number of stars in radial
bins located in the outermost cluster regions. Data plotted in Fig.~\ref{f4} show
the \ii,\rmi~ CMD for stars in radial bins located between $13\arcmin$ and
$2^{\circ}$ from the cluster center. The first three radial bins --panel a), b) 
and c-- are entirely included inside the CFHT dataset, while the last two 
--panel d) and e)-- are entirely located inside the SDSS dataset. The two 
solid lines display the acceptance region we defined in Fig.~\ref{f3}. 
The photometric precision in the outermost cluster
regions is clearly supported by the thin distribution of MS stars. Data plotted
in the three innermost radial bins indicate that MS stars are crucial to trace
the radial extent of candidate cluster stars. Moreover, the ratio between the 
number of cluster stars and the total number of stars is steadily decreasing 
when moving toward the outermost cluster regions. It decreases from almost 
$50\%$\ ($N_A/N_T$=0.49$\pm$0.02) for R$\sim 14.5\arcmin$ to slightly less than 
one third ($N_A/N_T$=0.30$\pm$0.01) for R$\sim 25\arcmin$. This radial distance 
appears to be a preliminary plausible lower limit for the truncation radius, 
and indeed the same ratio in the two outermost radial bins attains smaller 
constant values. Note that the main vertical sequence partially overlapping 
with the acceptance region is almost entirely made up of field stars.

\begin{figure}[!ht]
\begin{center}
\includegraphics[height=0.85\textheight,width=1.0\textwidth]{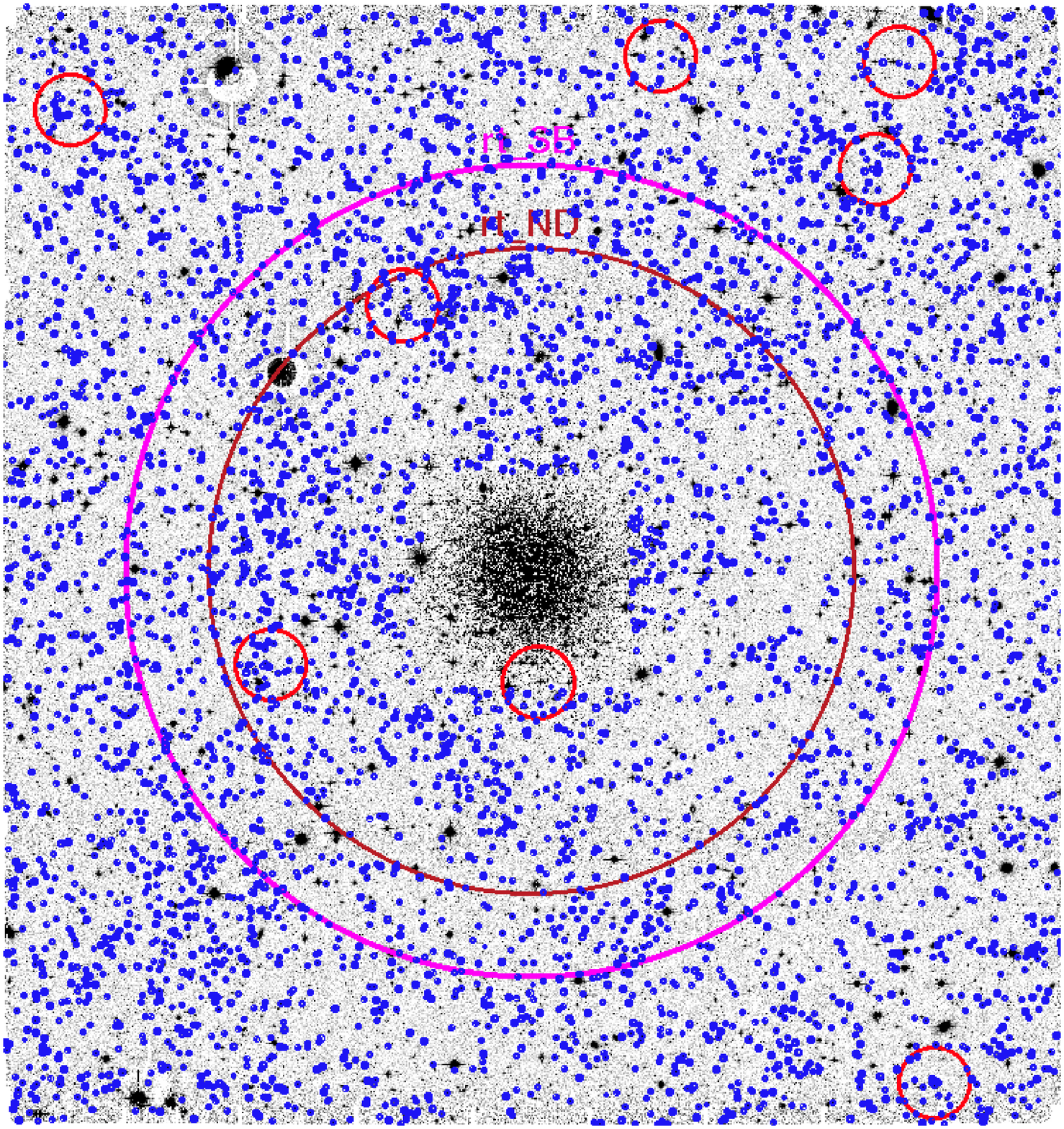}
\vspace*{-0.15truecm}
\caption{M92 {\it r}-band image of the CFHT dataset. The blue dots 
display the non-point like sources identified using Sextractor. The large 
purple and red circles display the Wilson truncation radius according to 
the SB and 
the ND profile. The small orange circles show the candidate galaxy clusters 
identified by Wen et al. (2009). The orientation is: North up and East left, 
the field of view is $1^{\circ}\times1^{\circ}$. 
\label{f5}
}
\end{center}
\end{figure}

To further constrain the plausibility of the above working hypothesis
concerning the radial extent of M92, we decided to investigate the
radial distribution of extragalactic sources. We adopted the entire
set of {\it r}-band images collected with MegaCam at CFHT and performed a new
independent photometry by using Sextractor (Bertin \& Arnouts 1996).
The non-point like sources were selected --following Evans et al. (2010)--
as the objects with a local point spread function 90\% enclosed counts
fraction larger than 1.4$\arcsec$. The adopted value was fixed by eye
inspection of the mean {\it r}-band image. The Fig.~\ref{f5} shows
the radial distribution of the non-point like sources (blue dots) over
the mean {\it r}-band image with apparent magnitudes between {\it r}$\sim$19
and  {\it r}$\sim$22 mag. To help the eye to identify the sky area covered
by M92, the red and the purple circles display the Wilson truncation radius 
based on both the number density  and the surface brightness profile
(see \S4). The smaller orange circles show the candidate galaxy
clusters identified from the SDSS DR6 (Wen et al. 2009).
A glance at data plotted in this figure shows that the candidate galaxy
clusters are located either close to the truncation radius or beyond it. 
The innermost candidate galaxy cluster appears a bit suspicious, since 
it is located in a cluster region with a high stellar density.

We selected the objects located inside the sky region covered by the five 
candidate galaxy clusters
beyond the truncation radius and plotted them in the CMD and we found that
a significant fraction ($\sim$65\%) of them are located outside the acceptance
region (blue triangles in panel c) of Fig.~\ref{f4}). This fraction agrees, 
within the errors, quite well with the ratio between accepted and total 
number of stars (see the discussion above in this section).

\begin{figure}[!ht]
\begin{center}
\includegraphics[height=0.8\textheight]{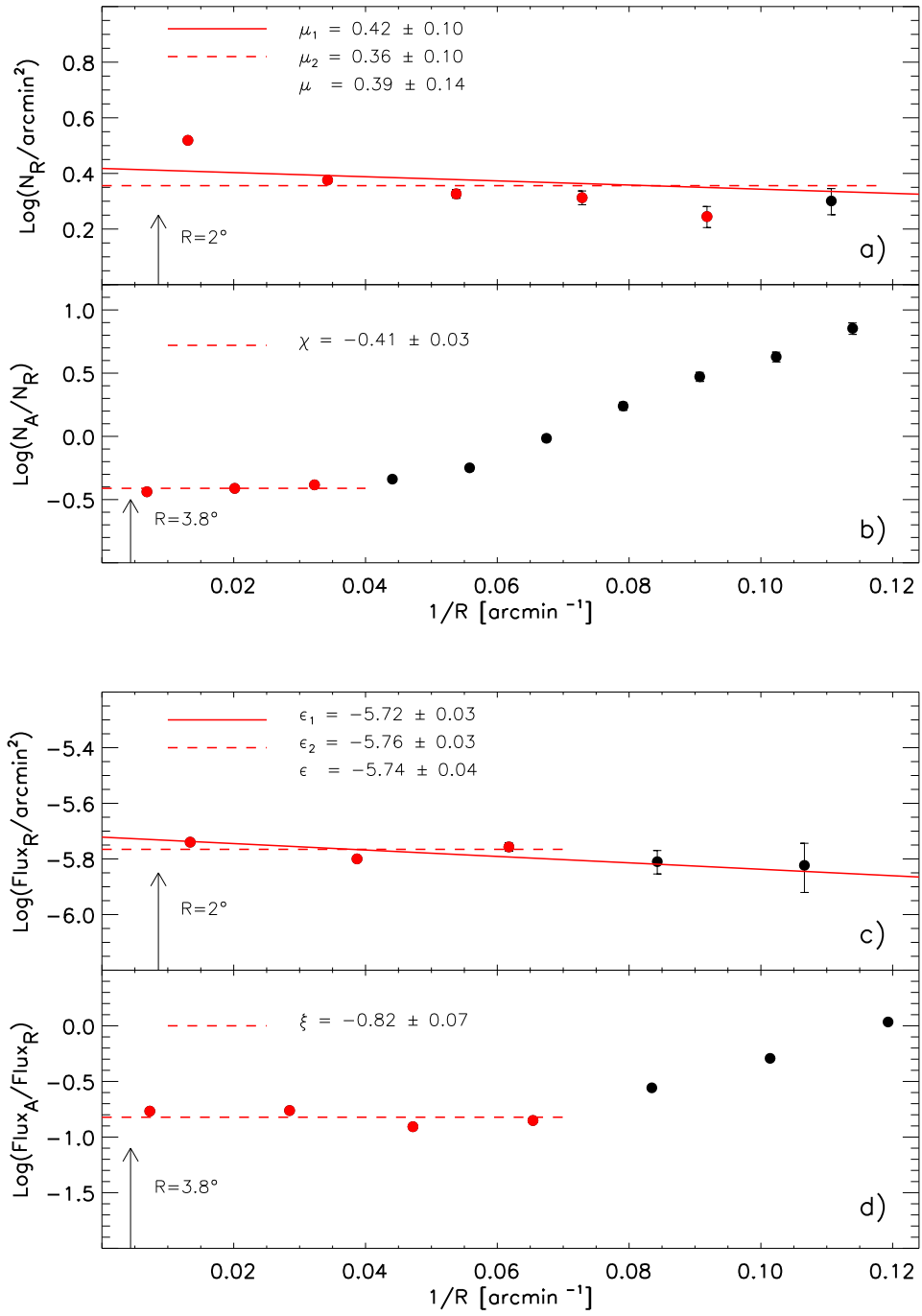}
\caption{Panels a) and b): logarithm of the number density of the {\it rejected} stars 
and of the ratio between the number of {\it accepted} and {\it rejected} --$N_A/N_R$-- 
stars as a function of the inverse of the radial distance. Panels c) and d): logarithm 
of the flux density of the {\it rejected} stars, and of the ratio between the flux of 
{\it accepted} and {\it rejected} --$Flux_A/Flux_R$-- stars. The vertical arrows 
display the edge of outermost annulus, the solid lines show the linear fits over 
the entire samples, the dashed lines indicate the mean of the outermost annuli 
(red points). See text for more details.
\label{f6}
}
\end{center}
\end{figure}

To remove spurious stars that were erroneously {\it accepted}, we used the 
method described by Walker et al. (2011) for the GC IC 4499. This method works 
particularly well for our datasets, due to the large sky area they cover. 
To estimate the density of the rejected stars, the sky area covered by CFHT 
($R\lesssim$30$\arcmin$) and by SDSS (0.5\deg $\lesssim R  \lesssim$2\deg) data 
was divided into concentric annuli. The star counts based on SDSS data were 
corrected to account for the non homogeneous coverage of this photometric 
catalog. 
Panel a) of Fig.~\ref{f6} shows the logarithmic surface density of these 
objects (number of rejected stars, $N_R$, per arcmin$^2$) as a function of the 
inverse of the radial distance. We performed a linear fit to the individual 
points and by extrapolating to infinite radial distance we found that the 
asymptotic value is $\mu_1=$0.42$\pm$0.10 [logarithmic number of stars per 
arcmin$^2$]. We also estimated the asymptotic value as the mean of the five 
outermost values, finding $\mu_2$=0.36$\pm$0.10. We adopted the mean of 
the above estimates: $\mu$=0.39$\pm$0.14. 

In panel b) of Fig.~\ref{f6} we plotted the logarithm of the ratio 
between the number of accepted stars and the number of rejected stars 
($N_A/N_R$). This ratio was estimated using the entire dataset. 
To estimate the mean asymptotic value, we adopted the average of the outermost three 
radial bins, obtaining $\chi$=--0.41$\pm$0.03. Eventually, by multiplying 
the number of rejected stars per arcmin$^2$ by $N_A/N_R$, we found the 
number of candidate field stars that were erroneously classified as 
candidate M92 stars. Notably, we found that the asymptotic number 
of spuriously accepted stars is 10$^\mu\times10^\chi\sim0.95$~star/arcmin$^2$. 
By subtracting this value from the number of the {\it accepted} 
stars per unit area, we obtained the final Count Catalog of the 
candidate M92 stars. This catalog was used to compute the Number 
Density (ND) radial profile. We divided the cluster into concentric annuli 
and we counted the number of stars per arcmin$^2$ that fall inside each 
region, obtaining a radial profile ranging from $R\sim1.5\arcsec$ out 
to $R\sim$2\deg. The error on each of the points is calculated as 
the square root of the number of stars, divided by the area of the 
annulus.

Following Walker et al. (2011), we measured the spurious stellar 
flux which affects the stellar luminosity of {\it accepted} stars. 
The logarithmic flux density of the rejected stars per arcmin$^2$ 
($Flux_R$) and the ratio between the flux of accepted and rejected 
stars ($Flux_A/Flux_R$) are plotted in panels c) and d) of Fig.~\ref{f6}. 
The logarithm of the surface flux density of the rejected stars approaches 
$\epsilon_1$=--5.72$\pm$0.03 [logarithmic star flux per arcmin$^2$] 
when extrapolated to infinite radial distance, 
whereas it is $\epsilon_2$=--5.76$\pm$0.03 when we use the mean value of 
the last three points. The mean of these values is $\epsilon$=--5.74$\pm$0.04, 
whereas the logarithm of the ratio between accepted and rejected stars is 
$\xi$=--0.82$\pm$0.07. In this case, the flux of spurious accepted stars is 
10$^\epsilon\times10^\xi\sim0.3~10^{-6}$ Flux/arcmin$^2$. By subtracting this last 
value from the {\it accepted} stellar flux density, we obtained an independent 
final Flux Catalog for the candidate M92 stars. This catalog was used to 
calculate the SB radial profile. As in the previous case 
we divided the cluster into annular regions, and the SB for each annulus was computed 
by adding the flux contribution of the stars located inside the annulus 
and by dividing for its area. 
Data plotted in panels b) and d) of Fig.~\ref{f6} show that the ratio  
between the number of accepted and rejected stars is more robust than 
the ratio between the flux of accepted and rejected stars, since the 
intrinsic dispersion of the former one is at least a factor of two smaller 
than the latter one. The difference is caused by the fact that the 
number ratio is rooted in the radial distribution of MS stars, while 
the flux ratio traces the radial distribution of bright evolved stars. 
To further constrain the role of bright stars in determining the radial 
slope of the SB, we calculated two more SB profiles by considering only 
stars fainter than a limiting magnitude of \ii=15 (SB-15) and \ii=17 
(SB-17) mag (see Sect.~\ref{Sect.4}). 

The error on the surface brightness profile was estimated by propagating
the error on individual measurements of star magnitudes. The intrinsic
photometric error, in the magnitude range adopted to estimate the surface
brightness profile, is typically of the order of a few hundredths of a
magnitude (see \S~2), because ground-based images were collected in good 
seeing conditions (DiC10) and we typically have more than ten images per band.
The same applies for ACS images adopted in the central regions. We also
calculated the error of the absolute photometric zero-points.
Following DiC10, we calibrated the CFHT photometric catalog by using the 
local standards by Clem et al.\,(2007). The ACS and the SDSS photometric 
catalogs were also transformed into the same photometric systems by using 
the new local standards. We ended up with a mean calibration error in the 
$i$-band of 0.02$\pm$0.04 mag (Di Cecco 2009). This error was eventually 
summed in quadrature with the intrinsic photometric error.

\begin{figure}[!ht]
\begin{center}
\includegraphics[height=0.75\textheight,width=0.6\textwidth]{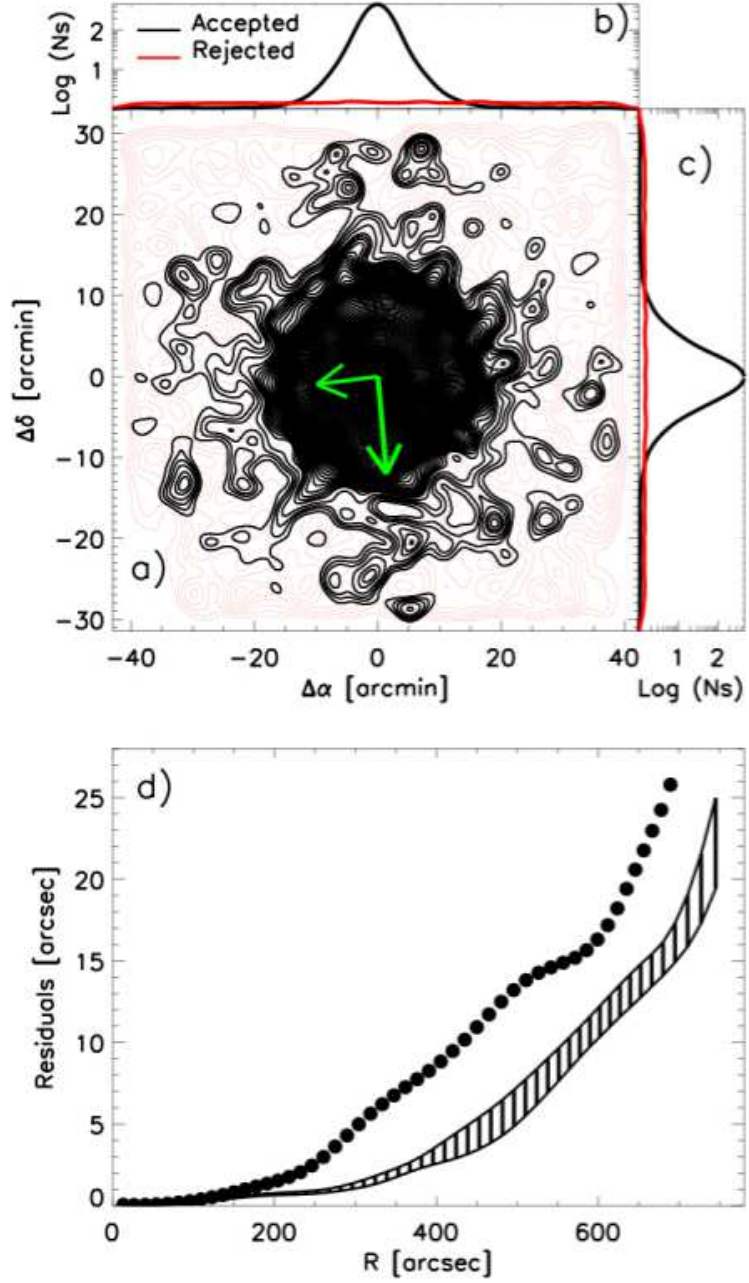}
\caption{Panel a): contour levels for candidate M92 (black) and field (red) stars 
projected onto the sky. The long green arrow marks the direction of the Galactic 
Center, the short one the M92 proper motion. Panels b) and c): projected logarithmic 
distribution (marginal) along the horizontal and the vertical axis for candidate 
M92 (black) and field (red) stars. Panel d): the black filled circles show the 
residuals of the fits to the contour levels plotted in panel a) with circles of 
variable radius as a function of the radial distance. The vertically hatched 
area shows the results of simulations (see text for more details).
\label{f7}
}
\end{center}
\end{figure}

We evaluated the symmetry of the ND as a function of the radial distance 
by using the Count Catalog. 
To avoid possible systematic uncertainties in the radial distribution,
the symmetry of the ND was estimated on the basis of ACS and CFHT data.
The SDSS data were neglected since they do not uniformly cover the
area of the sky around the cluster center. The conclusions concerning
the departure from circular symmetry of the radial distribution for
distances between $6\arcmin$ and $30\arcmin$ are not affected by the
inclusion of the SDSS dataset.
We computed the contour levels of the candidate M92 and field stars 
(black and red lines in panel a) of Fig.~\ref{f7}). 
%
The contour levels become less circular symmetric when moving toward the outermost 
regions, and for radial distances between 6$\arcmin$ and 10$\arcmin$ the 
stellar density decreases by almost one order of magnitude (see top panels 
of Fig.~\ref{f7}).
The contour levels become asymmetric at a distance of $\sim$13$\arcmin$
(see panel a) of Fig.~\ref{f7}), which is almost equivalent to the truncation 
radius of M92 (see column 5 in Table~1) available in the literature. Outside 
this region the distribution of candidate cluster stars becomes clumpy. 
Panels b) and c) of Fig.~\ref{f7} show the projected linear density (marginal) 
along the axes of the candidate M92 and field stars (black and red lines). 
By inspecting these panels, we notice that the detected sharp decrease in 
density is associated with the radial distance at which the surface density 
of the candidate M92 stars becomes smaller than that of the candidate field 
stars.

%
To trace in detail the departure of the contour levels from circular symmetry,
we performed a fit of each contour with a circle. The center of the circles is
identical to the center of the contours --i.e. the cluster center-- and 
the radius of the circles is the fitting parameter. Then, we computed the 
residuals in arcsec between individual contours and best fitting circles. 
The residuals were estimated from the very center of the cluster out to a radial 
distance of $R\sim800\arcsec$. 

The residuals plotted in panel d) of Fig.~\ref{f7} 
show that the innermost contour levels display symmetric radial distributions,
and indeed the residuals attain vanishing values out to $R\sim$3$\arcmin$.
%
At larger radial distances, the asymmetry increases out to
$R\sim$9$\arcmin$--10$\arcmin$ ($R\sim$500--600$\arcsec$),
where the residuals show a shoulder clearly connected with the density 
drop detected in the contour plot. 
At even larger distances the contours become more asymmetric 
(residuals~$\sim$~26\arcsec ~for distances of $R\sim13\arcmin$); 
the fit in the outermost regions fails to converge 
because of the large asymmetries. The increase in asymmetry that 
we found in the outer 
regions could be the consequence of the fluctuations associated with the 
decrease in density. To validate this working hypothesis we performed a 
series of simulations by using the observed density profile to compute 
synthetic GCs; the radial distribution of the synthetic GCs was required to be 
symmetric. 
We applied to these GCs the same procedure to evaluate the contour levels 
and the same fit with circles of variable radius. The vertical hatched area 
plotted in panel d) of Fig.~\ref{f7} marks the residuals calculated for the 
synthetic GCs. The comparison between this area and the plotted points 
indicates that the asymmetry in the real cluster is at least 3$\sigma$ larger 
than in the synthetic clusters. To characterize further the nature of the 
asymmetries in the contour levels, we show two green arrows, in the top panel 
of Fig.~\ref{f7}, indicating the direction of the Galactic center (long arrow) 
and of the M92 proper motion (short arrow) according to Dinescu et al. (1999). 
We found no clear correlation between these directions and the clumpy distribution 
of candidate M92 stars at large radial distances.  

Evidence of a clumpy stellar distribution in the outskirts of M92 was also
present in the stellar density maps provided by Testa et al. (2000, see their
Fig.~6), by Lee et al. (2003, see their Fig.~12) and by Jordi \& Grebel (2010,
see their Fig.~17). The above results support the findings by Lee et al. (2003),
 concerning the marginal evidence of an elongation of outermost clumpy stars in 
the direction orthogonal to the direction of the Galactic center.

\section{Dynamical models and fits}
\label{Sect.4}

\begin{figure}[!ht]
\begin{center}
\includegraphics[height=0.75\textheight]{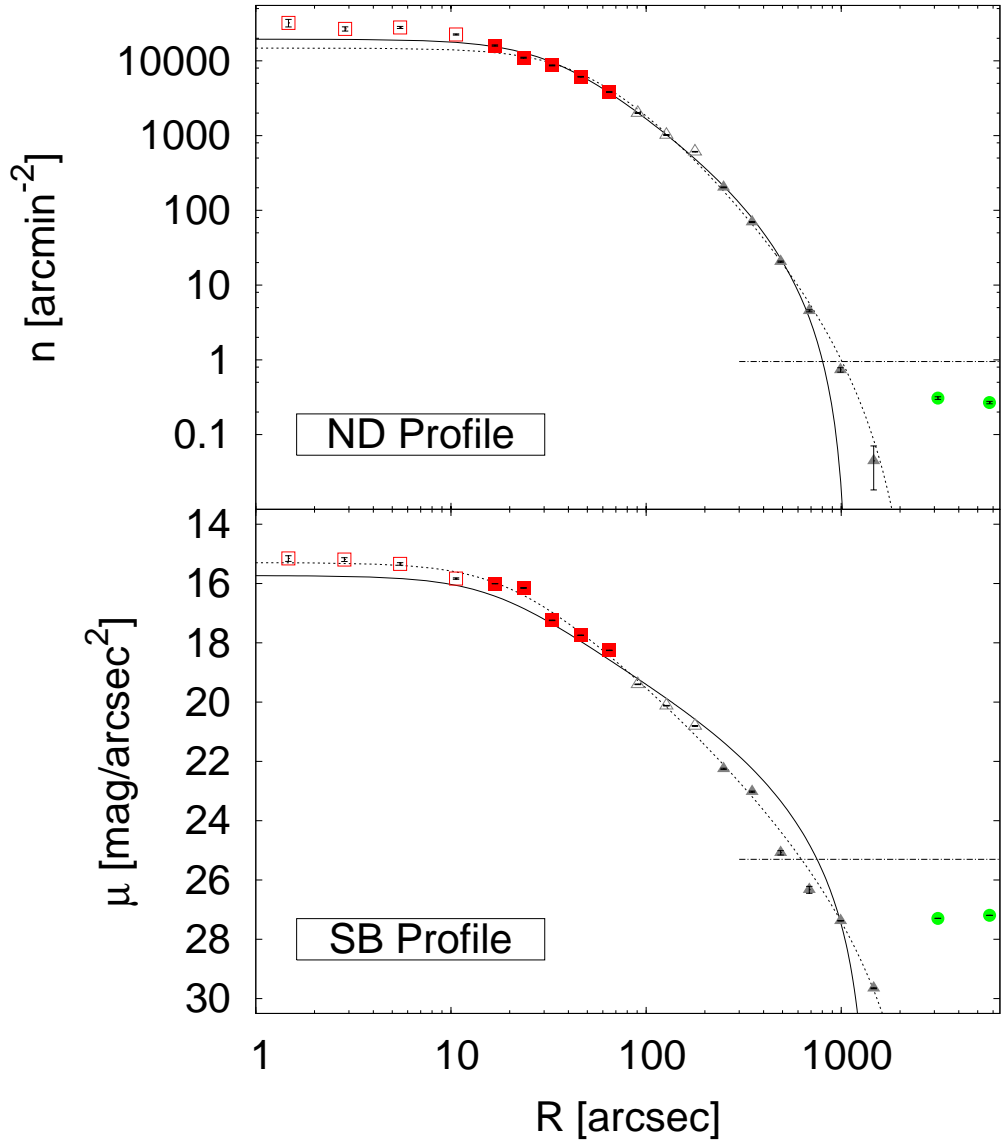}
\caption{Fits by King and Wilson spherical models to the ND (top) and SB (bottom) 
profiles. Solid lines correspond to the King-model fits, dotted lines to Wilson-model 
fits; the horizontal dashed line shows the background level; errors are shown as vertical 
error bars. Red squares indicate data obtained from pointing $\gamma$, grey triangles 
those from the CFHT dataset, and green circles those from the SDSS. Empty symbols 
mark regions for which a completeness correction was applied.} \label{f8}
\end{center}
\end{figure}

\begin{figure}[!ht]
\begin{center}
\includegraphics[height=0.75\textheight]{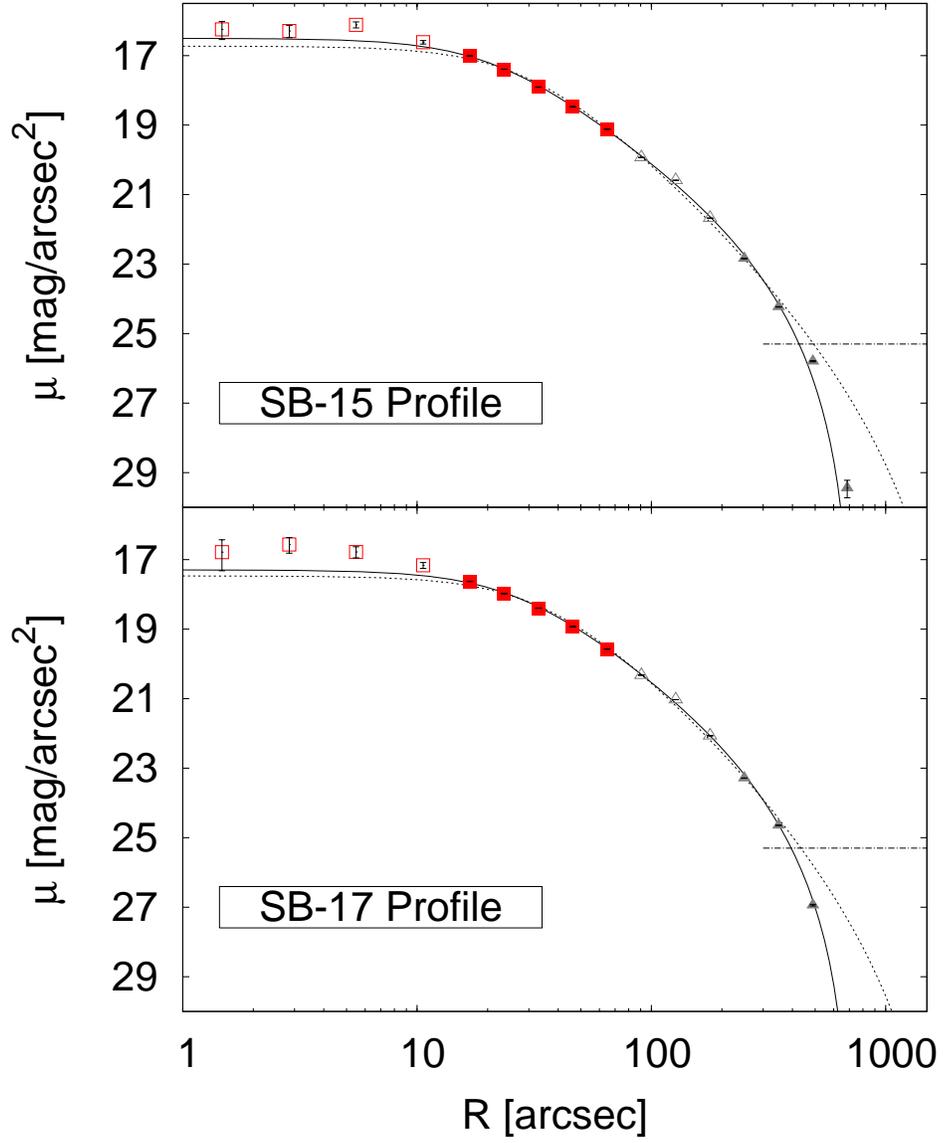}
\caption{Fits by King and Wilson spherical models to the SB-15 (top) and the 
SB-17 (bottom) profiles. Solid lines correspond to the King-model fits, dotted 
lines to Wilson-model fits; the horizontal dashed line shows the background level. 
Data-points are indicated as in Fig.~\ref{f8}, and errors are shown as vertical 
error bars. \label{f9}
}
\end{center}
\end{figure}

We carried out fits of dynamical models to the observed radial profiles.
We considered the King (1966) and the Wilson (1975) spherical and isotropic
dynamical models, defined by the following distribution functions:
\begin{equation}
f_{\mathrm{K}}=A\left( e^{-aE}-e^{-aE_0}\right)\ \  E\leq E_0 \ ,
\label{King_f}
\end{equation}
\begin{equation}
f_{\mathrm{W}}=A\left\lbrace
e^{-aE}-e^{-aE_0}\left[1-a(E-E_0)\right]\right\rbrace \ \  E\leq E_0 .
\label{Wilson_f}
\end{equation}
The quantity $E$ is the specific star energy $E=v^2/2+\Phi(r)$,
where $\Phi(r)$ is the mean-field gravitational potential, to be determined
from the Poisson equation. Both distribution functions vanish for energies
larger than the threshold energy $E_0$, corresponding to stars to be considered
as unbound. The energy truncation can be translated into a truncation radius, 
$r_t$, which indicates the boundary of the system. For each family of models, 
the constants $A$, $E_0$, and $a$
in Eqs.~(\ref{King_f}) and (\ref{Wilson_f}) define two dimensional scales
(a typical radius and a typical mass or velocity) and one dimensionless
parameter, the central depth of the potential well (related to the concentration
parameter). We recall that fitting by a one-component dynamical model 
assumes that the underline stellar populations are distributed homogeneously. 
To identify the best-fit model we adopted the procedure described by 
Zocchi et al. (2012).
The results are shown in Figs.~\ref{f8} and \ref{f9}; the values of the 
relevant parameters of the best-fit models are listed in the upper part 
of Table~\ref{table1}.

The fit to the ND profile is shown in the top panel of Fig.~\ref{f8}.
Both King and Wilson models underestimate the central ND profile, 
failing with respect to the four innermost points. 
A quantitative interpretation of this discrepancy remains 
unavailable, but the problem is likely to be related to the failure 
of the assumptions at the basis of a one-component dynamical model 
in the central regions. 
Concerning the King models, the good
agreement with observations that can be found in the middle part of the
profile breaks down around 700$\arcsec$, where the profile approaches the
background level (0.95~star/arcsec$^2$).
In the case of the Wilson models, instead, only the two outermost points
are discrepant, and the model fits the data out to a distance
greater than 1000$\arcsec$. Note that the two outermost points 
are likely to be affected by errors in the subtraction of 
background stars.
The bottom panel of Fig.~\ref{f8} shows the fits to the SB profile.
When compared to the King best-fit model, the Wilson best-fit model 
provides a more adequate overall description, not only in relation 
to the outermost points, as in the ND profile, but also in the central 
part of the profile.
The satisfactory performance of the Wilson models indicates that the
observations can be explained by means of a less abrupt truncation radius,
with no need to introduce extra-tidal halos. In this case, the two outermost
points were not taken into account to calculate the best-fit parameters,  
since they are expected to be even more affected by errors in the 
subtraction of background stars, compared to the 
corresponding points in the ND profile. In the figure, the background 
level (25.30 mag/arcsec$^2$) is indicated as a horizontal dashed line.

Surprisingly, even if the ND and the SB profiles come from the same set of
observations, the best-fit parameters determined by the fits are significantly
different. We argue that this behavior is due to the fact that each profile
represents a different aspect of the density distribution of the cluster.
On the one hand, the ND profile, derived by considering the radial distribution 
of both luminous and faint stars, is dominated by the MS stars, which greatly 
outnumber evolved (RGB, HB) stars (Castellani et al. 2007). On the other hand, 
the SB profile is heavily affected by the presence of the brighter RGB stars. The
difference in the best-fit models reflects the intrinsic difference in the radial
distribution of the stellar tracer that determines each profile. This behavior
should be interpreted as a signature of mass 
segregation\footnote{The occurrence of mass segregation in M92 was also 
suggested by Andreuzzi et al. (2000).}. Indeed, the evolved slightly more 
massive stars\footnote{Note that hot HB stars are less massive than
MSTO stars, but they are a minor fraction of evolved cluster stars.} appear to be
more centrally concentrated compared to stars with lower masses (see the
values of the concentration parameter $c$ in Table~\ref{table1}).

This interpretation is confirmed by an additional test. We carried out the same
fitting procedure on the SB-15 and the SB-17 profiles (as defined in Sect.~3). 
The resulting parameters
are listed in the third and fourth rows of Table~\ref{table1}. The fits
are shown in the top and in the bottom panel of Fig.~\ref{f9}, respectively.
Inspecting the values of the best-fit parameters, it appears that by
eliminating the brightest stars, the profiles tend to approach the ND profile.
Indeed, the values of the concentration parameter and of the radial scale
--$r_s$-- follow a monotonic trend from the SB profile to the SB-15, to the
SB-17 and finally to the ND profile.
Interestingly enough, for the SB-15 and SB-17 profiles, the King models 
reproduce the data better than the Wilson models, in contrast to what we found 
for the previously described ND and SB profiles. A reason for this can be found 
by comparing the radial extent of the different profiles.
The three surface brightness profiles are shaped by the fact that the cut 
of candidate cluster stars brighter than a limiting magnitude causes a 
decrease in the radial extent and a more abrupt truncation of the profile.
Indeed, the outermost radial point for the SB-15 profile is located at
$R\sim700\arcsec$, and for the SB-17 profile is located at
$R\sim500\arcsec$. The larger radial extent of the SB profile is the
consequence of a few bright giants that keep the SB profile well
above the background level.

For comparison, the middle part of Table~\ref{table1} lists 
the most recent best-fit parameters published for M92. Note that in 
only a few cases the Wilson parameters are available. The acronyms 
in the first column identify the papers in which the results have been 
published.

\begin{figure}[!ht]
\begin{center}
\includegraphics[height=0.75\textheight]{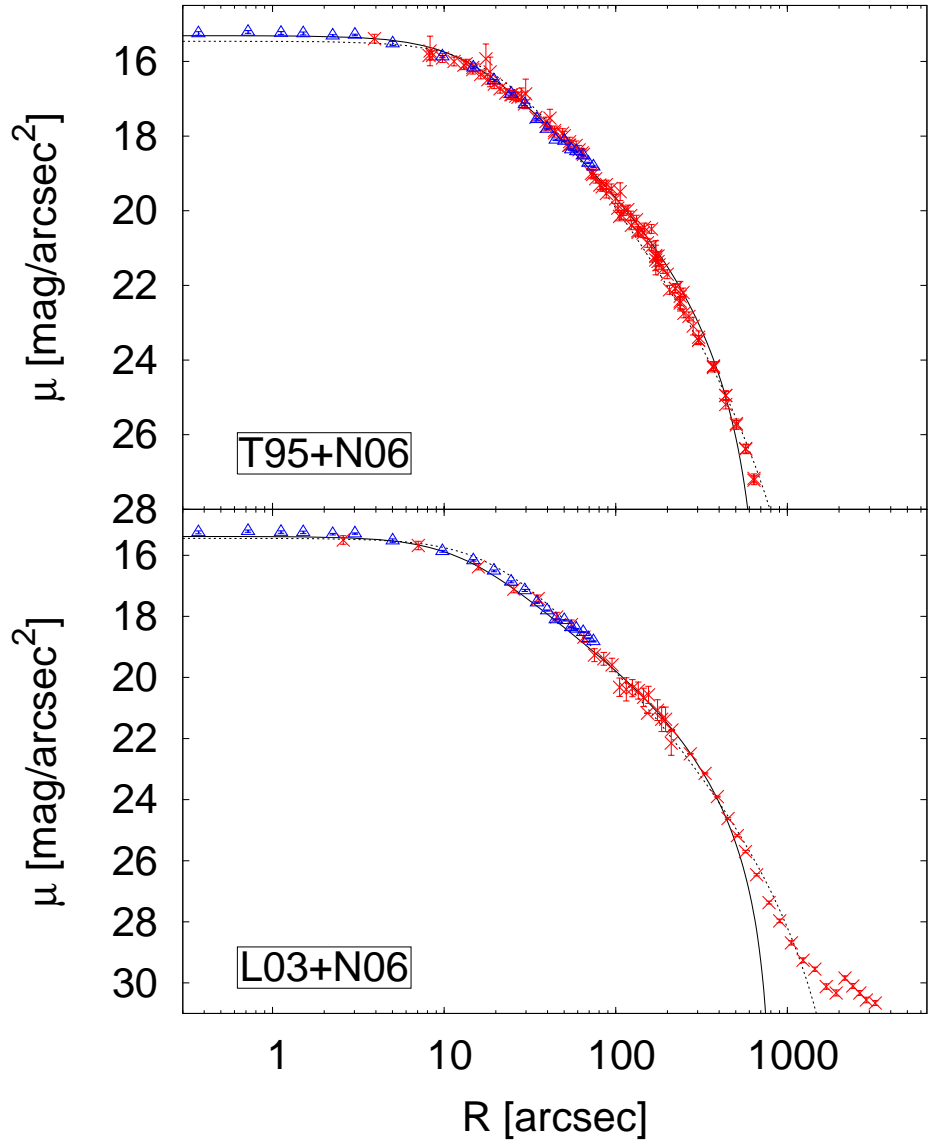}
\caption{Fits by King and Wilson spherical models to the composite T95+N06 (top) and 
to the L03+N06 (bottom) profiles. Solid lines correspond to the King-model fits, dotted 
lines to Wilson-model fits; errors are shown as vertical error bars. Blue triangles 
indicate data from N06, red crosses those from other sources.\label{f10}
}
\end{center}
\end{figure}

The lower part of Table~\ref{table1} lists the results of the fits 
to the other two\footnote{Another surface brightness profile was 
presented by JG10, but the dataset is not publically available.}  
available surface brightness profiles in the literature, the one by T95, 
and the one by L03. We decided to add to both datasets the surface 
brightness data by Noyola et al. 2006 (N06), which are relative to the 
innermost region of the cluster. The results of these fits are shown in 
Fig.~\ref{f10}. For the T95+N06 profile, the models provide an equally 
good fit to the data; we note that in this case the outermost point in 
the profile is located at a radial distance R$<700$$\arcsec$. For the 
L03+N06 profile, the good agreement with observations that is seen in 
the inner and middle parts of the profile breaks down around 600$\arcsec$ 
for the King model, and at more than 1000$\arcsec$ for the Wilson model. 
We recall that a similar behavior was also found for the fit to the 
ND profile previously described. 
In the last profile the outermost points are not reproduced 
by any of the considered models.

The values of the parameters listed in Table~\ref{table1} are not 
consistent with each other, within the errors. In conclusion, we believe 
that a proper comparison of the values of the best-fit parameters found 
by fitting models to different (ND, SB) profiles requires that we take 
into account the role of the different stellar tracers in determining 
their shape, which makes standard one-component dynamical models questionable.

\section{Discussion and final remarks}

We studied the radial distribution of stars of the globular cluster M92 by
using ground-based (MegaCam at CFHT, SDSS) and space (ACS on HST) data.

The contour levels, based on star count data, are symmetric in the
innermost regions, and exhibit an increasing asymmetry for radial distances
between 3$\arcmin$ and 9$\arcmin$--10$\arcmin$. For distances larger than
$\sim$13$\arcmin$ the stellar distribution becomes clumpy. The contour levels
do not exhibit a preferred orientation in space.
We calculated two independent radial profiles, to describe the distribution 
of stars in the cluster, the number density (ND) and the surface brightness
(SB) profile. To calculate these profiles, we subtracted the background contamination
with two independent methods. We performed fits of spherical King and Wilson models to
the above profiles. Wilson models appear to reproduce, better than King models, the
behavior of the outermost regions of the cluster with no need of extra-tidal halos.
Interestingly, for the ND profile, both models significantly underestimate the
observations in the innermost regions.

We also found that the best fit to the ND and to the SB profile are provided by two
different models for the two families, even though the profiles are derived from 
the same datasets. We argue that this difference is caused by a difference in the 
radial distribution of the stellar tracers that characterize the two observed 
profiles.  The ND profile traces the radial distribution of MS stars, whereas the
SB profile that of bright evolved (RGB, HB) stars. This conclusion is supported
also by the results of a test that has been carried out on two additional profiles,
calculated by considering only stars fainter than a given magnitude.

Hopefully, a thorough discussion of the behavior in this and other clusters
should determine which of the various profiles considered is best suited for 
a study in terms of one-component models. To our knowledge, this is the 
first investigation in which independent estimates of both ND and SB profiles 
are provided, starting from the same set of data, and compared.

In this context a key role can be played by the new generation of wide field imagers
that are available at the 4-8m class telescopes (Dark Energy Camera and Survey
at the CTIO 4m Blanco telescope, Mohr et al. 2012; Hyper SuprimeCam at
SUBARU\footnote{http://www.naoj.org/Projects/HSC/index.html}). In a single
pointing they can cover the entire extent of a large number of GCs and with modest
exposure times will allow us to perform homogeneous and accurate photometry several
magnitudes fainter than the MSTO.

\acknowledgments
It is a real pleasure to thank an anonymous referee for his/her positive 
opinion concerning the content and the cut of our paper and his pertinent 
suggestions that helped us to improve its readibility.  
This work was partially supported by PRIN--INAF 2011, `Tracing the formation
and evolution of the Galactic halo with {\sl VST}' (P.I. M. Marconi), by
PRIN--MIUR (2010LY5N2T) `Chemical and dynamical evolution of the
Milky Way and Local Group galaxies' (P.I.: F. Matteucci), by the IAC
(grants, 310394) and by the Education and Science Ministry of Spain 
(grants AYA2010-16717).
One of us, A. Di Cecco, thanks the warm hospitality of the Dominion 
Astrophysical Observatory for the stay during which part of this work has been 
done; G.~Bono thanks ESO for support as a science visitor. This publication
makes use of data products from VizieR \citep{ochse00} and from the Two Micron
All Sky Survey, which is a joint project of the University of Massachusetts and
the Infrared Processing and Analysis Center/California Institute of Technology,
funded by the National Aeronautics and Space Administration and the National
Science Foundation. We also used the NASA/IPAC Extragalactic Database (NED) which is
operated by the Jet Propulsion Laboratory, California Institute of Technology, under
contract with the National Aeronautics and Space Administration.


\clearpage
\begin{center}
\begin{deluxetable}{l ccccc ccccc}
\tabletypesize{\scriptsize}
\tablewidth{0pt}
\tablecaption{Best-fit parameters. \label{table1}}
\tablehead{
Profile& \multicolumn{4}{c}{King models}& \multicolumn{4}{c}{Spherical Wilson models}\\
\colhead{}&
\colhead{$\Psi$}&
\colhead{c}&
\colhead{$r_{\mathrm{s}}$}&
\colhead{$r_t$}&
\colhead{$\Psi$}&
\colhead{c}&\colhead{$r_{\mathrm{s}}$}&
\colhead{$r_t$}&}
\startdata
ND      &  6.91$\pm$0.02 & 1.50$\pm$0.01 & 34.25$\pm$0.38 & 18.11$\pm$0.44 &  5.84$\pm$0.02 & 1.73$\pm$0.01 & 48.03$\pm$0.42 & 42.63$\pm$1.19 \\
SB      &  8.40$\pm$0.01 & 1.95$\pm$0.00 & 15.22$\pm$0.02 & 22.80$\pm$0.18 &  6.65$\pm$0.01 & 2.14$\pm$0.00 & 21.08$\pm$0.05 & 48.80$\pm$0.85 \\
        &       &      &       &         &       &      &       &       \\
SB-15   &  7.20$\pm$0.01 & 1.59$\pm$0.00 & 19.43$\pm$0.05 & 12.56$\pm$0.10 &  6.52$\pm$0.01 & 2.06$\pm$0.00 & 24.33$\pm$0.06 & 46.67$\pm$0.42 \\
SB-17   &  6.95$\pm$0.01 & 1.51$\pm$0.00 & 23.17$\pm$0.06 & 12.63$\pm$0.12 &  6.29$\pm$0.01 & 1.94$\pm$0.00 & 29.04$\pm$0.06 & 41.97$\pm$0.53 \\
        &       &      &       &       &       &      &       &       \\
T95     &  7.92 & 1.81 & 23.67 & 15.20 &       &      &       &       \\
T00     &       &      &       & 12.33 &       &      &       &       \\
L03     &  8    & 1.83 & 12.42 & 14.00 &       &      &       &       \\
MLvdM05 &  7.5  & 1.68 & 16.15 & 12.88 &  5.9  & 1.75 & 26.51 & 24.85 \\
JG10    &  6.93 & 1.51 & 23.37 & 12.55 &       &      &       &       \\        
        &       &      &       &       &       &      &       &       \\
T95+N06 &  7.54 & 1.69 & 14.72 & 12.07 &  6.34 & 1.96 & 19.33 & 29.58 \\
L03+N06 &  7.84 & 1.78 & 13.46 & 13.67 &  6.61 & 2.12 & 18.94 & 41.72 \\
\enddata
\tablecomments{For each model, we list the dimensionless parameter $\Psi$, the concentration $c$, 
the scale radius $r_s$ (arcsec), and the truncation radius $r_t$ (arcmin). The different profiles 
are identified by the label in the first column. The cases indicated as T95+N06 and L03+N06 
refer to the fits we performed on composite profiles, obtained by combining the profiles from 
T95 and L03 with the profile from Noyola et al. (2006, [N06]), which covers the innermost 
regions of the cluster.}
\end{deluxetable}
\end{center}

\clearpage 

\end{document}